\documentclass[twocolumn,showpacs,preprintnumbers,prb,fleqn,eps]{revtex4}
\usepackage{epsfig}
\usepackage{graphicx}
\usepackage{amsfonts}

\begin{document}

\title{Local electronic properties in nanoscale systems}
\author{Andre M. C. Souza$^{1,2}$ and Hans Herrmann$^{3,4}$}

\affiliation{$^{1}$Institut f\"{u}r Computerphysik, Universit\"{a}t
Stuttgart, Pfaffenwaldring 27, 70569 Stuttgart, Germany}

\affiliation{$^{2}$Departamento de Fisica, Universidade Federal de
Sergipe, 49100-000 Sao Cristovao-SE, Brazil}

\affiliation{$^{3}$Computational Physics, IfB, ETH H\"{o}nggerberg,
HIF E 12, CH-8093 Z\"{u}rich, Switzerland}

\affiliation{$^{4}$Departamento de F\'{i}sica, Universidade Federal
do Cear\'{a}, 60451-970 Fortaleza-CE, Brazil}

\date{\today}
\begin{abstract}
The local electronic structure on nanoscale chains is investigated
theoretically. We propose a mechanism to explain the even-odd
oscillation of length distribution of atom chains. We study the
spatial peak structure as obtained by scanning tunneling microscopy
(STM) constant-current topography as a function of the
electron-electron interaction, band filling and temperature. The
site-dependent magnetic moment is also examined.
\end{abstract}
 \maketitle

\section{INTRODUCTION}

The recent development of techniques capable of constructing
structures on the nanoscale, with dimensions that are intermediate
in size between isolated atoms/molecules and bulk materials, has
opened up numerous possibilities for constructing new devices. This
perspective raises the importance about novel and improved physical,
chemical, mechanical, and biological properties of nanoscale atomic
clusters. In particular, a relatively recent focus has been to study
the effects of the electronic quantum confinement within finite
length chains \cite{ves,cutb,qwell,d1,mi,mn}. With the help of
scanning tunneling microscopy (STM) the formation of quantized
chain-localized states in the pseudogap of the substrate bulk band
in chains at semiconductor or insulator surfaces has been observed
\cite{cutb}.

The investigation of the electronic properties of nanoscale atomic
clusters is particularly interesting. The electronic properties
suffer strong effects due to the finite length of chains. The
breaking of translational symmetry creates electronic end states.
The electronic quantum confinement in finite length segments changes
the density of states within the chains. The conductivity along the
chain shows spatial variations in the electronic density of
states\cite{cutb,qwell,d1}. The direct observation of the local
electronic structure on a nanoscale atomic chain by STM stresses the
importance of exploring these effects theoretically.

We can observe that while some STM measurements can be explained on
the basis of non-interacting electrons \cite{cutb,qwell,d1}, others
instead suggest strong interaction between electrons \cite{mn}
indicating the possibility of a description using the Hubbard model.
The significance of the results for extremely small systems without
a systematic analysis of the observed effects for larger systems
would not have been appropriate some years ago, when the
investigation of new nanoscale devices was known only as a
theoretical possibility. Since the experimental skills to manipulate
the nanostructures, this study has become important, not only to
obtain new insight regarding properties of theoretical models, but
certainly also to better understand physical properties in this
class of systems.

In this paper we study theoretically the local behavior of the
physical quantities of open chains on the nanoscale. We report on
site-dependent properties of the electronic quantum confinement
within finite length chains. This theory allows for an
interpretation of experimental results on finite-size effects. We
studied two different aspects of these systems. First, considering
only the electronic structure, we introduce a model that captures
qualitatively the experimental chain lengths distribution. For this
case, we analyzed the internal structure of atom chains by examining
the site-dependent occupation number along the axis of the chain for
one electron. Secondly, we studied the electron-electron interaction
by using the Hubbard model to understand the main effects of the
electronic quantum confinement within finite length chains.

The organization of this paper is as follows. Our investigation of
the chain lengths distribution using the tight binding approach is
presented in Sec. II. The results for local occupation numbers and
for the site-dependent magnetic moment as a function of the
electron-electron interaction, band filling and temperature on
electronic quantum confinement within finite length chains are
presented in Sec. III. Our conclusions are presented in Sec. IV.

\begin{figure}[!ht]
\includegraphics[width=85mm]{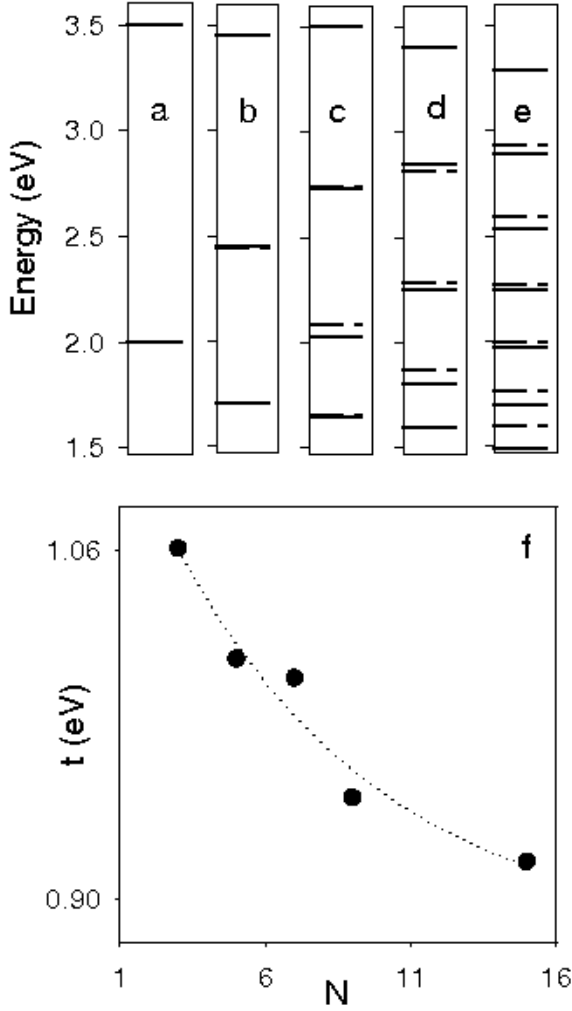}
\caption{Experimental (full line), extracted from Ref. \cite{cutb}
and theoretical (dotted line) eigenenergies for chains of (a) three,
(b) five, (c) seven, (d) nine and (e) fifteen atoms. (e) The electron hopping
strength $t$ as function of the chain length.} \label{fi1}
\end{figure}

\section{TIGHT BINDING APPROACH}
\subsection{Introduction}
Fitting of experimental wave vectors of nanoscale atomic clusters
have shown a one-dimensional (1D) free electron band dispersion
relation\cite{cutb,crain2}. The 1D quantum potential well has been
proposed to describe the confinement of the electrons and the local
conductivity determined by the superposition of wave functions
\cite{cutb,qwell,qwell2}. In contrast to a superposition of wave
functions of the 1D potential well \cite{qwell,qwell2,cutb}, the
experimental results are described here by the site-dependent
occupation number along of the chain. We have used the tight binding
Hamiltonian
\begin{equation}
{\cal H}_{0} = -t \sum_{i\alpha} (c_{i\alpha}^{\dag}c_{i+1\alpha} +
hc), \label{tb}
\end{equation}
where $c_{i\alpha}^{\dag} (c_{i\alpha})$ are the creation
(annihilation) operators for the electrons of spin $\alpha$ at site
$i$ and $t$ is the nearest neighbor hopping integral representing
the overlap of electron wave functions. To gain a quantitative
picture of the on site-dependence on a chain of $N$ atoms, we have
computed the local occupation number
\begin{equation}
n_{m}(i)=\langle \Psi_{m} | \hat{n}_{i}  | \Psi_{m} \rangle /N,
\end{equation}
where $\hat{n}_{i}=\hat{n}_{i\uparrow}+\hat{n}_{i\downarrow}$,
$\hat{n}_{i\alpha}=c_{i\alpha}^{\dag}c_{i\alpha}$ and $| \Psi_{m}
\rangle $ is the $m$th eigenvector with energy $E_m$ of ${\cal
H}_{0}$.

As illustration of our approach, we analyze the experimental data of
the eigenenergies for chains of three, five, seven, nine and fifteen
atoms extracted of Ref. \cite{cutb}. We have fitted the data taking
the experimental ground-state energies and considering $t$, in the
tight binding Hamiltonian of Eq. 1, conveniently to obtain the
experimental bandwidths. The theoretical and experimental energy
spectra are shown in Figs. \ref{fi1}a-e.  The experimental data are
reproduced. The electron hopping strength as function of the chain
length can be observed in Fig. \ref{fi1}f. Note that, as the chain
length increases the electronic hopping decreases. The change of
hopping amplitudes should be caused by changes of the underlying
atom chains.

In the following will investigate the question of the length
distribution of atom chains.

\begin{figure}
\psfig{figure=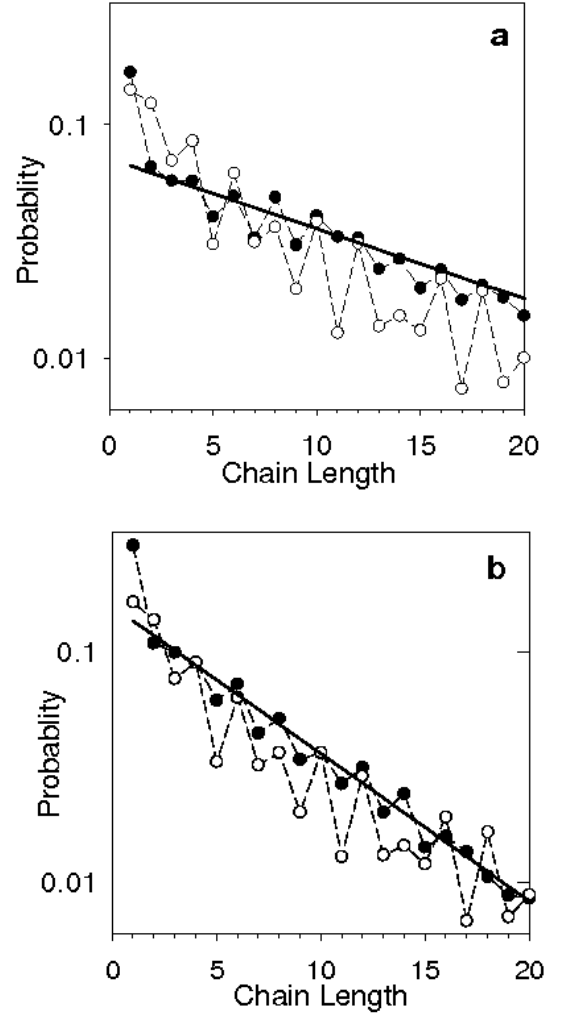,width=80mm,angle=0} \caption{Experimental
(full circle), extracted from Ref. \cite{leng1} and theoretical
(open circle) distribution of chain lengths for $t/(k_{B}T)=0.5$ and
(a) low defect density ($\rho =0.0625$) and (b) medium defect
density ($\rho =0.125$).} \label{fi2}
\end{figure}

\subsection{Length distribution}
Recent results on the distribution of chain segments of gold
deposited on a Si surface reveal a relation between the chain length
and the cohesive energy \cite{leng1}. The size distribution is
characterized by a strong peak for a length of one atom and even-odd
oscillations, where even chain lengths are favored over odd lengths.
We have obtained a direct relation between this experimental result
and the local electronic density of the tight binding model on a
finite open chain. We find that on some sites the probability of
finding electrons in the excited energy levels is zero. We also
observe that these sites correspond to positions that divide the
chain in $m$ subchains of equal length. The delocalized electronic
structure determines the stability of the atom chain. The cohesive
energies are a consequence of the local electronic structure
\cite{leng1,leng2}. In particular, we observe that the sites with
zero probability to find electrons on odd chains are not binding,
and result in a larger probability for breaking chains. To address
this problem we propose a simple model. First, we consider the
distribution of chain lengths for a completely random distribution
of defects to be $\rho (1-\rho )^N$, where $\rho$ is the defect
density \cite{leng1}. Since chains are fabricated at temperature $T
\sim 1000 ^{\circ }$C and considering $t \sim 1$ eV, it is relevant
to evaluate the canonical average for any quantity  $x$ as
\begin{equation}
\langle x \rangle \equiv \sum_{i} x_{i} exp(-E_{i}/k_{B}T)/Z,
\end{equation}
where $Z$ is the partition function. Next, we assume that there are
quantum states breaking a chain of length $m$ into $g$ pieces of
length $N$. It follows that $g \equiv g_{m,N}=(m+1)/(N+1)$ and we
can write the distribution of chain lengths as
\begin{equation}
P(N)= A \sum_{m=1}^{\infty } \rho (1-\rho )^m \langle q(m,N) \rangle
,
\end{equation}
where $A$ is a normalization constant,
$q_{i}(m,N)=g_{k,N}\delta_{kN}$,  $k=k(m,i)$ is a non-trivial
function obtained numerically, and $\delta_{kN}$ is the Kronecker's
delta.

Fig. \ref{fi2} shows the experimental data, extracted from Ref.
\cite{leng1}, and our theoretical chain lengths distribution for low
and medium defect densities. The even-odd oscillations are clearly
visible. Notably, the present model captures qualitatively the
experimental chain lengths distribution, considering only the
electronic structure effects. The quantitative disagreement between
theory and experiment is due to the coupling of the electronic
structure with the substrate. This coupling was observed in Ref.
\cite{leng1} using photoemission measurements of the electronic
scattering vectors at the Fermi surface of the surface states.

Furthermore, a mechanism for the change in chain length obtained by
comparing STM images of the same sample region taken at different
voltage \cite{crain2} is also confirmed by our results. Taking
different voltages in STM, the system can fall into an energy level
that singles out to a position, which divides the chain in
subchains. An additional nearest-neighbor interaction, reflecting
the end states in the chains, was used to describe this effect in
Ref. \cite{crain2} and has been already observed in other STM
topography data \cite{aurev,y}.

\begin{figure}
\psfig{figure=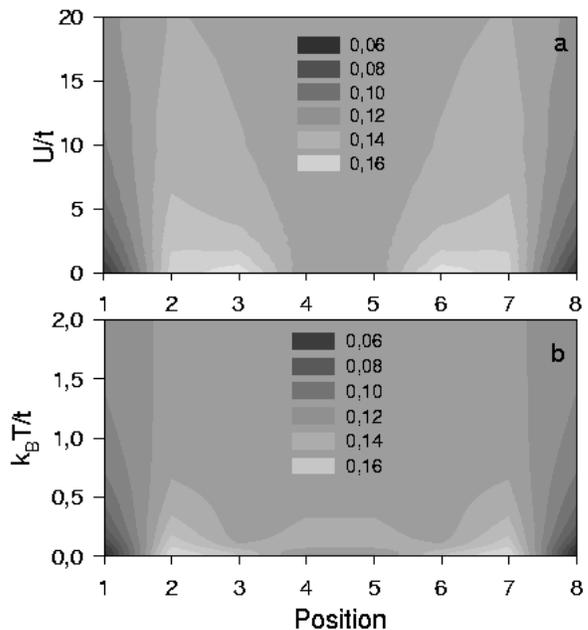,width=80mm,angle=0} \caption{Values of
$n_{0}(i)$ for a chain of $8$ sites in the quarter-filled band (a)
versus $U/t$ for temperature $T=0$ and (b) versus $T$ for $U/t=8$.}
\label{fi3}
\end{figure}

\section{Electron-electron interaction}

In addition to the tight binding model, containing one electron, it
is possible to vary the band filling of chain structures
\cite{cutb}. A strong electronic interaction was found for small
chains of Mn on CuN, and the STM data were consistent with the
Heisenberg model results \cite{mn}. In this case, the effect of
electron-electron interaction is particularly important.

The possibility of explaining some STM measurements by the tight
binding model\cite{cutb,qwell,d1} and others measurements by the
Heisenberg model\cite{mn} indicates that we can use a description
using the Hubbard model. The 1D Hubbard model is defined by the
Hamiltonian
\begin{equation}
{\cal H} = {\cal H}_{0} + U \sum_{i} \hat{n}_{i\uparrow}
\hat{n}_{i\downarrow} . \label{hamil}
\end{equation}
where $U$ is the on-site Coulomb (electron-electron) interaction.

Solutions for the 1D Hubbard model are known since the sixties
\cite{lieb}. The local behavior of the physical quantities is
however not completely known. Only partial informations are
available \cite{1d2,1d3,x}. For example, the spatial dependence of
the occupation number and of the magnetization was calculated for
the 1D Hubbard model with open boundary conditions at zero
temperature combining numerical computations from density-matrix
renormalization group and Bethe-Ansatz methods \cite{1d4}. It is
important to observe that Friedel oscillations for the density and
the magnetization in open Hubbard chains have been obtained in
previous works (see \cite{gia} and references therein), however, the
focus of these studies has been to analyze the physical properties
of systems having macroscopic size.

We perform exact calculations of physical quantities of 1D systems
having $N$ atoms described by the Hubbard model. We use the standard
direct diagonalization method \cite{shiba,meu} and impose open
boundary conditions in order to break the translational symmetry.

We have found in a half-filled band for $U=0$ that the occupation
number of the ground state is site-{\em independent}.  For $U >0$
this result is valid for all states. This is relevant in order to
evaluate the thermodynamic quantity $\langle n(i) \rangle $. We
obtain in a half-filled band that $\langle n(i) \rangle =1/N$ for
all sites $i$, temperatures $T$ and couplings $U>0$. This is a
valuable result because the sites are not equivalent due to the use
of open boundary conditions. For other than half-filled bands, we
observe a site-dependence of $n_{m}(i)$. For a quarter-filled
Hubbard band we have found a rich dependence on $U$, $m$, $i$ and
$T$. Fig. \ref{fi3}a presents the topography of $n_{0}(i)$ versus
$U/t$ for a chain of $8$ sites in the quarter-filled band for the
temperature $T=0$. At low values of $U/t$, the intermediate atoms
have a greater value than average and the end and central atoms have
smaller values. Here, the probability of finding electrons on the
intermediate atoms is higher. The site-dependence decreases if the
electron-electron interaction increases. For $U/t \rightarrow
\infty$ the site-dependence disappears and for all $i$ we find
$n_{0}(i)=1/8$. We can alternatively fix $U/t$ and vary the
temperature in order to cover the other energy states. Fig.
\ref{fi3}b shows this for $U/t=8$ (other values of $U/t$ give
similar results). The effect of increasing the temperature is
equivalent to an increase of  $U/t$. Large temperature destroys the
site-dependence of the charges at the chain. It is important to
comment that while this statement appears true for the local
occupation number, there are other quantities, such as intersite
spin correlations, that may not follow this rule.

\begin{figure}
\psfig{figure=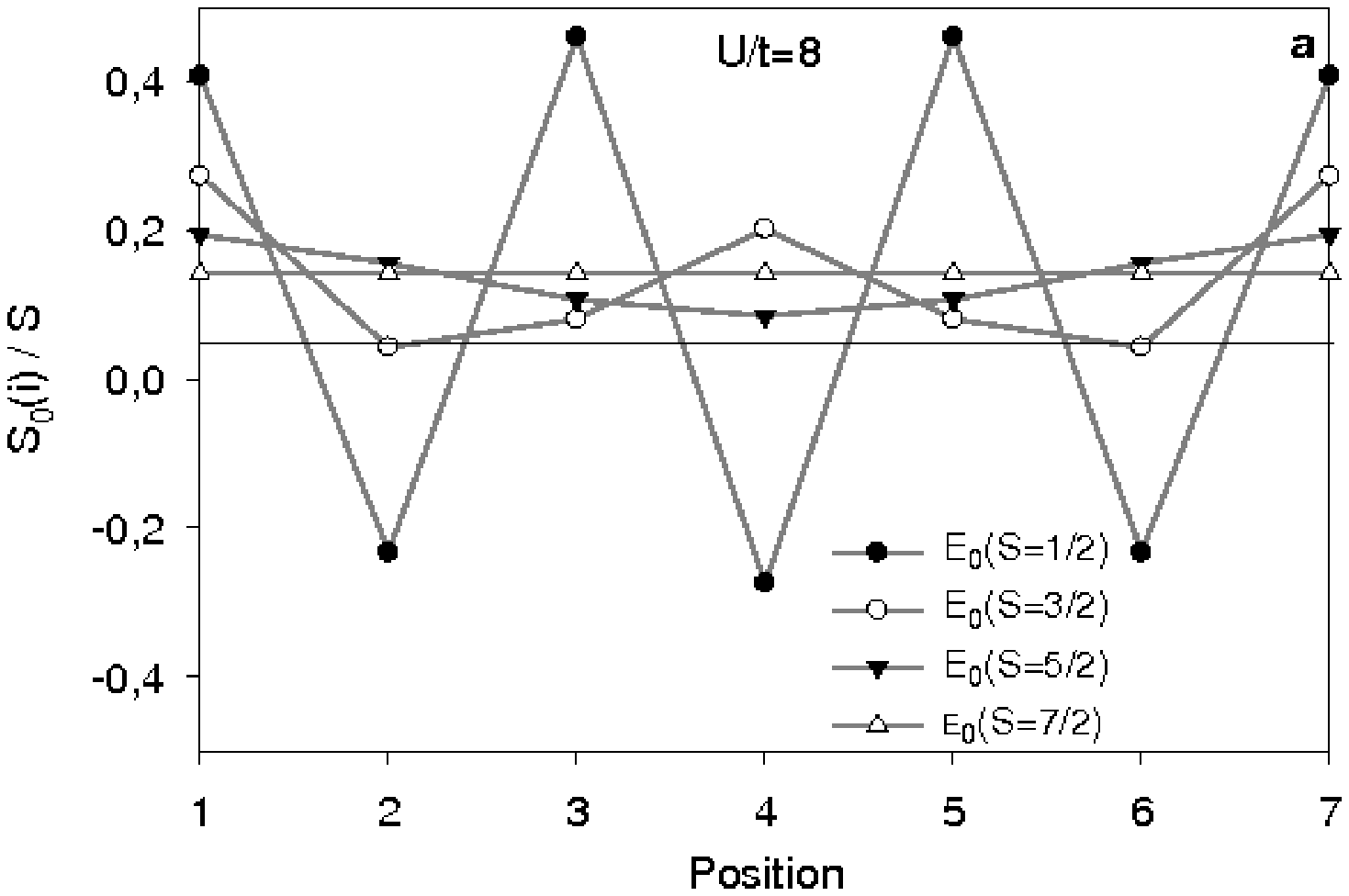,width=80mm}
\psfig{figure=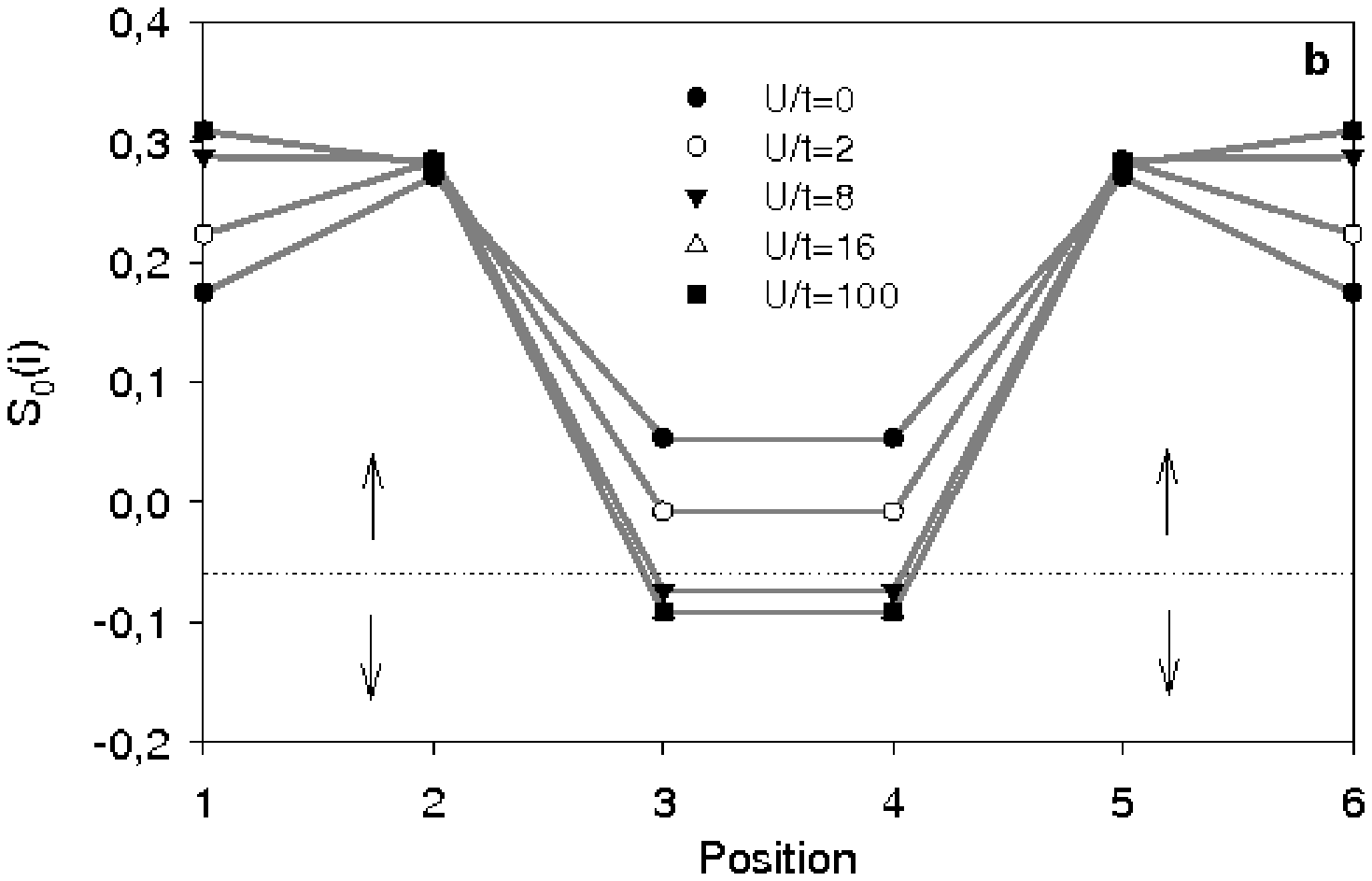,width=80mm} \caption{(a) Site-dependent spin
$S_{0}(i)/S$ versus the position $i$ for a chain of $7$ sites, for a
half-filled band, $U/t=8$ and different total spin $S=\sum_{i}
S_{0}(i)$. (b) $S_{0}(i)$ for a chain of $6$ sites for a
quarter-filled band versus position $i$ for temperature $T=0$ and
$U/t=0,2,8,16$ and $50$.} \label{fi4}
\end{figure}

For the magnetic properties of finite chains, first-principle
calculations have shown that for small chains the spin moment
depends on position and cluster length.
For example, for Co chains on Pt(111), the spin moments at the end atoms are
higher than those at the central atoms and the spin moment of the central atom
was found to decrease if the chain size increases \cite{t}
Results of the present study are illustrated in Fig. \ref{fi4}.
We have studied the local spin number
\begin{equation}
S_{m}(i)=\langle \Psi_{m} | \hat{S}^{z}_{i}  | \Psi_{m} \rangle,
\end{equation}
where $\hat{S}^{z}_{i} = ( \hat{n}_{i\uparrow}-\hat{n}_{i\downarrow} )/2$
at site $i$.

The local spin number $S_{0}(i)$ versus the position $i$ for a chain
of $7$ sites, for the half-filled band and temperature $T=0$, can be
observed for different total spins $S=\sum_{i} S_{0}(i)$ and
$U/t=8$. For $S=1/2$ the alternate sign of $S_{0}(i)$ for odd and
even position gives evidence for an antiferromagnetic structure. The
cases $S > 1/2$ have a ferromagnetic structure, since all $S_{0}(i)$
have the same sign. Site-independence is found for $S=7/2$. On the
other hand, when $S$ is between the minimal and maximal values we
find a site-dependence for $S_{0}(i)$, whose amplitude still
alternates but not sufficiently to change its sign. For $S=5/2$ we
obtain that $S_{0}(i)$ is higher at the end atoms than at the
central atoms as has been found for Co chains on Pt(111) \cite{t}.
An important aspect is the influence of the band filling and the
cluster length. In particular, we have found that for the case of
even total electron number ($n_{m}=\sum_{i} n_{m}(i)$) the
$S_{m}(i)$ is site-independent in all $m$ states. $\langle S(i)
\rangle =0$ for all sites $i$, temperatures $T$ and couplings $U$.
Considering $N$ even, using that $\langle n(i) \rangle =1/N$ in the
half-filled band and $\langle S(i) \rangle =0$, we have that the
canonical ensemble averages are site-independent, like in
mean-field, and $\langle n_{\uparrow}(i) \rangle =\langle
n_{\downarrow}(i) \rangle =1/(2N)$ indicating that the open finite
even chain is paramagnetic in the half-filled band. It is important
to mention that we find site-independence although the system has no
translational symmetry. The situation for odd finite chains is
different. In this case $\langle n(i) \rangle = 1/N$,
 but $\langle S(i) \rangle$ has a complex site-dependence.
For the ground state, at $U=0$ we obtain a structure in which $S_{0}(i)=1/(N+1)$ and
on the nearest-neighbor sites $ S_{0}(i)=0$ ($\sum_{i} S_{0}(i)=S_{0}^{z}=1/2$).
At $U > 0$ we observe an antiferromagnetic structure.

Finally, we explore the dependence of the site-dependent spin on the coupling $U$.
Fig. \ref{fi4}b shows $S_{0}(i)$ for the quarter-filled band
versus $i$ for a chain of $6$ atoms and typical $U/t$ values.
The ground state has $S=1/2$. Here, we show
the case $S^{z}=1/2$. For low $U/t$ all $S_{0}(i)$ have the same sign favoring a
ferromagnetic order and the higher $S_{0}(i)$ values are at intermediate atoms and
the smaller at the central atoms. Increasing $U/t$ the central atoms assume
negative values for the spin and the higher $S_{0}(i)$ shift to the end atoms
following a $\uparrow \uparrow \downarrow \downarrow \uparrow \uparrow$ magnetic
stripe structure.

The control and manipulation of the spin rather than the charge of
electrons \cite{lu} in STM could show the very rich behavior of the
local magnetic properties presented here.  It would be of interest
to explore the possibility of obtaining experimental results in the
context of magnetic force microscopy (MFM) devices \cite{mfm}. This
would generate images that are sensitive to the spin polarization of
the electrons. The site-dependent magnetic moment (spin) shown here
could be confirmed by these experiments.

\section{CONCLUSION}
We have shown that the spatial peak structure of STM
constant-current topography can be explained in terms of the
site-dependent occupation number. We interpret the even-odd
oscillation of the length distribution of atom chains. We also have
studied the spatial structure of the electronic charge and spin on
finite chains, involving band filling, temperature and
electron-electron interaction and we examined the site-dependent
magnetic moment (spin) indicating properties that may be confirmed
by the MFM device. This work provides a simple novel theoretical
explanation for the electronic confinement in atomic chains and
opens up a way for a more detailed analysis of the recent STM
measurements.

\section*{ACKNOWLEDGMENTS}
The help of E. Parteli is gratefully acknowledged. This work was
supported by CNPq (Brazil) and DAAD(Germany) and the Max Planck
prize.

\end{document}